\documentclass[aip,apl,amsmath,amssymb,reprint,groupeaddress,showpacs]{revtex4}

\usepackage{verbatim}
\usepackage{hyperref}
\usepackage{graphicx}
\usepackage{dcolumn}
\usepackage{bm}
\usepackage{sidecap}
\usepackage{epstopdf}
\begin{document}

\title{Transparent gradient index lens for underwater sound based on phase advance}%

\author{Theodore P. Martin}
\email{theodore.martin@nrl.navy.mil}
\affiliation{U.S. Naval Research Laboratory, Code 7160, Washington, DC 20375, USA}
\date{\today}

\author{Christina J. Naify}%
\author{Elizabeth A. Skerritt}%
\author{Christopher N. Layman}
\author{Michael Nicholas} 
\author{David C. Calvo}
\author{Gregory J. Orris}%
\affiliation{U.S. Naval Research Laboratory, Code 7160, Washington, DC 20375, USA}

\author{Daniel Torrent}%
\author{Jos\'e S\'anchez-Dehesa}%
\affiliation{Department of Electronic Engineering, Wave Phenomena Group, Universidad Politecnica de Valencia, C/ Camino de Vera s7n, E-46002 Valencia, Spain}


\begin{abstract}

Spatial gradients in refractive index have been used extensively in acoustic metamaterial applications to control wave propagation through phase delay.  This study reports the design and experimental realization of an acoustic gradient index lens using a sonic crystal lattice that is impedance matched to water over a broad bandwidth.  In contrast to previous designs, the underlying lattice features refractive indices that are lower than the water background, which facilitates propagation control based on a phase advance as opposed to a delay.  The index gradient is achieved by varying the filling fraction of hollow, air-filled aluminum tubes that individually exhibit a higher sound speed than water and matched impedance.  Acoustic focusing is observed over a broad bandwidth of frequencies in the homogenization limit of the lattice, with intensity magnifications in excess of 7~dB.  An anisotropic lattice design facilitates a flat-faceted geometry with low backscattering at 18~dB below the incident sound pressure level.  Three dimensional Rayleigh-Sommerfeld integration that accounts for the anisotropic refraction is used to accurately predict the experimentally measured focal patterns. 

\end{abstract}

\maketitle

Metamaterial lattices composed of sub-wavelength scattering components have been used increasingly in recent years to control the propagation of both electromagnetic and acoustic waves through two- and three-dimensions. One primary goal of acoustic metamaterial design has been to achieve effective fluid, or ``metafluid,'' material properties that minimize shear coupling and propagation.  The coupling between a fluid-born acoustic wave and a composite elastic material at off-normal incidence results in mode-mixing of dilatation and shear modes that complicates the interaction.  For underwater applications requiring strong acoustic coupling, compliant materials such as rubbers have been a standard material used for coupling and encapsulation due to their relatively close impedance match with water and low shear moduli.  Traditional compliant materials often have larger densities and/or lower dilatation moduli than water, making them useful for applications requiring low relative sound speed.  For example, a sonic crystal lattice of rubber cylinders with a gradient filling fraction was recently used to achieve an underwater omnidirectional absorption coating~\cite{Naify2014APL}, which requires a radially decreasing sound speed.~\cite{Li2011APL,Climente2012APL}

Metafluids with complementary material properties compared to rubbers, e.g.~high relative sound speeds with matched impedance, have been more difficult to achieve in underwater applications because they require high stiffness but low relative density.  Sonic crystal lattices feature metafluid functionality in the homogenization limit of the lattice, including shear decoupling and broadband performance, and hence they have the potential to expand the range of realizable metafluid material properties to include high sound speeds.  Thin-walled, hollow elastic shells have been recently proposed as high-sound speed sonic crystal components in water.~\cite{Martin2012PRB}  By carefully tuning the wall thickness of the shell, an impedance-matched condition can be obtained assuming the shell material has higher impedance than the background fluid. Additionally, by using shell elements which are individually impedance matched, it has been shown that any phononic crystal configuration of the elements will also be impedance matched in the homogenization limit of the crystal lattice.~\cite{Martin2012PRB}  Given that local variations in the element filling fraction create a variation in refractive index, acoustically transparent devices with tunable wave-guiding capability should be achievable using hollow shell elements.  The concept of hollow shell elements has also been extended to include structural components inside the shells to improve control over the material properties.\cite{Titovich2014JASA}  

Here we report the design and experimental demonstration of an acoustically transparent gradient index (GRIN) lens that focuses sound by \emph{advancing} the phase of waves propagating through an aqueous background.  GRIN geometries have been utilized for a broad range of metamaterial applications including scattering reduction~\cite{Zhang2010PRL,Yin2012APL}, wave focusing~\cite{Torrent2007NJP,Climente2010APL,Martin2010APL,Peng2010APL,Sanchis2010APL,Zigoneanu2011PRB,Lin2012JAP,Chang2012NJP,Hladky2013APL} and bending~\cite{Ren2010APL,Layman2011APL,Li2011APL,Climente2012APL,Naify2014APL}.  In contrast to previous lens designs, we present a lens composed of impedance-matched, hollow shell elements with sound speeds higher than water.  The higher sound speed enables the phase of propagating waves to be advanced beyond what is possible in the background propagation medium, which is important to a number of metamaterial applications (for a review, see Refs.~\citenum{Maldovan2013Nat,Kadic2013RPP}).  The lens is intended as a proof of concept to demonstrate that both transparency and broadband wave-guiding functionality can be achieved simultaneously using a lattice of non-compliant, low refractive index scattering elements.   

The GRIN lens is constructed using cylindrical, air-filled aluminum tubes arranged in a lattice that function as a broadband, uniform effective medium at wavelengths $\lambda$ larger than the lattice spacing $a$ ($\lambda>4a$).   With a wall thickness of $1/20$th the diameter, each individual cylindrical tube scatters sound as an impedance-matched effective fluid cylinder~\cite{Martin2012PRB} resulting in negligible backscattering over the lattice as a whole.  An anisotropic lattice spacing is used to vary the filling fraction similar to what was proposed by Lin, Tittmann, and Huang~\cite{Lin2012JAP}.  The anisotropic design simplifies the lens construction by using a single cylindrical tube geometry and features a flat lens facet.  While the objective of many metafluid applications is to achieve anisotropy in the wave speed, the transparency of the lattice results in an approximately isotropic scattering configuration despite the underlying anisotropy of the design.  Our design goal is to tune the anisotropy to produce a convex refraction gradient that focuses sound while simultaneously minimizing lens aberration.

\begin{figure}[p]
\begin{center}
\includegraphics[width=11cm] {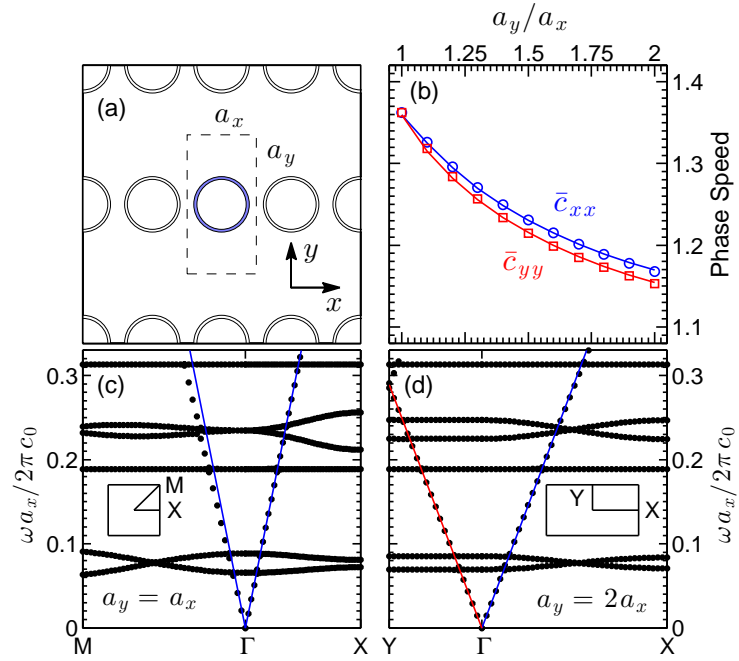}
\caption{(a) Schematic unit cell of the anisotropic lattice with $a_{y}=2a_{x}$. (b) Phase speeds in the homogenization limit of the lattice plotted as a function of anisotropy ratio $a_{y}/a_{x}$.  Lines are fits to the data based on Ref.~[\citenum{Torrent2008NJPb}]. (c) Acoustic band structure calculated using the Finite Element Method (FEM) for an isotropic lattice with tube outer radius $R=0.4a_{x}$.  Blue lines plot the acoustic sound speed in the $\Gamma$X direction. (d) Acoustic band structure calculated for an anisotropic lattice with tube outer radius $R=0.4a_{x}$.  Blue and red lines plot the acoustic sound speed in the $\Gamma$X and $\Gamma$Y directions respectively.  Insets show the directions of the bands in $k$-space.}
\label{fig:1}
\end{center}
\end{figure}

Figure~\ref{fig:1}(a) shows the rectangular unit cell of the cylinder lattice.  A variable refractive index can be achieved by lengthening one of the lattice constants $a_y$ compared to the lattice constant $a_x$ in the orthogonal direction. Phase speeds $\bar{c}_{xx}$ and $\bar{c}_{yy}$ in the primary Cartesian directions are shown in Fig.~\ref{fig:1}(b) as a function of the anisotropy ratio $a_{y}/a_{x}$ for a fixed aluminum tube outer radius $R=0.4a_{x}$.  The double index on the phase speeds delineates components in an anisotropic tensor.  Phase speeds $\bar{c}$ with an over-bar are normalized to the water background, which is assumed to have density $\rho_{b}=1000$~kg$/$m$^3$ and sound speed $c_{b}=1480$~m$/$s.  The phase speeds are derived from the longitudinal dispersion bands of the acoustic band structure calculated for each anisotropy ratio. The anisotropy in phase speed can also be derived by considering multiple scattering effects in the lattice.~\cite{Torrent2008NJPb} 

Examples of the acoustic band structure are shown in Fig.~\ref{fig:1}(c,d).  Blue and red lines, with slopes corresponding to the phase speeds in Fig.~\ref{fig:1}(b), demonstrate that the longitudinal bands have linear dispersion up to an upper cutoff frequency $\omega a_{x}/2\pi c_{b}\simeq 0.3$. The cutoff frequency is constrained by the wrapping of the longitudinal band at the Brillioun zone boundary in the $\Gamma Y$ direction for the highest anisotropy ratio considered.  The longitudinal band in the $\Gamma M$ direction in Fig.~\ref{fig:1}(c) has a slowness that deviates slightly from circularity due to the underlying cubic symmetry. In addition to the longitudinal bands, narrow resonance bands arising from the core-shell architecture of the unit cell are predicted at various frequencies.\cite{Titovich2014JASA}   While we do not observe evidence of coupling to these bands in our measurements, the resonant bands should not, in general, be overlooked depending on the lattice geometry.  

\begin{figure}[p]
\begin{center}
\includegraphics[width=13cm] {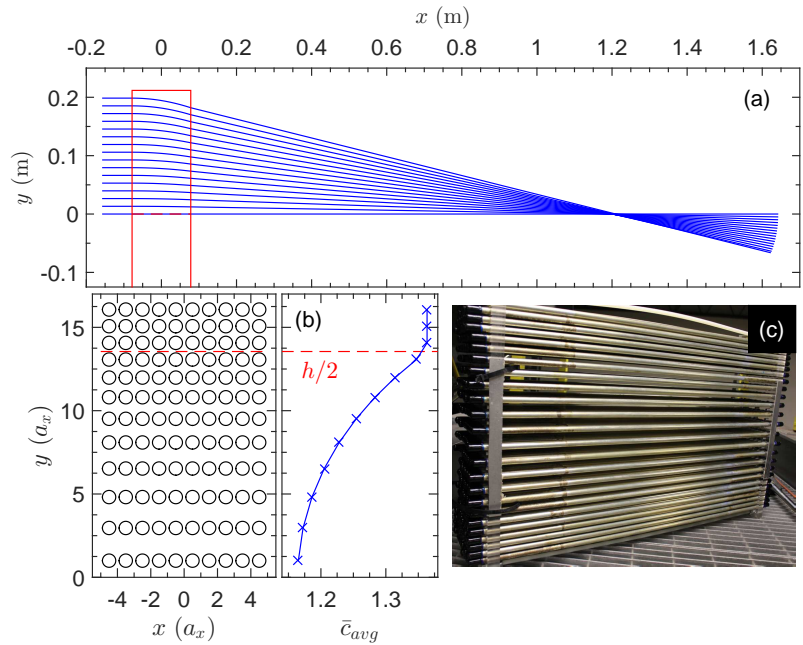}
\caption{(a) Ray paths through the lens (blue lines), which originate from a plane wave at normal incidence, are calculated using an eikonal approximation in the $xy$-plane; the red box indicates the location of the lens.  Only rays incident on the upper half-plane of the lens are shown.  (b) Schematic showing the locations of the lattice sites in the upper half-plane of the lens; the resulting sound speed profile is plotted to the right. The lattice sites are symmetric about $y=0$.  (c) Photograph of the assembled lens composed of 1~m long, hollow aluminum cylinders arranged in the desired lattice pattern.}
\label{fig:2}
\end{center}
\end{figure}

Our lens design consists of a constant lattice spacing in the $x$-direction, which produces a flat lens facet, while a prescribed lattice anisotropy in the $y$-direction produces the convex GRIN profile [see Fig.~\ref{fig:2}(b)].  The primary impact of increasing the anisotropy ratio is to decrease the effective sound speed irrespective of propagation direction.  As is evident in Fig.~\ref{fig:1}(b), the difference between $\bar{c}_{xx}$ and $\bar{c}_{yy}$ is small even at $a_{y}/a_{x}=2$, resulting in approximately isotropic acoustic transport over the entire Brillioun zone.  Therefore, an approximate, isotropic sound speed $\bar{c}_{avg}=\left(\bar{c}_{xx}+\bar{c}_{yy}\right)/2$ was used to initially guide the design of the GRIN geometry. A number of GRIN profiles that result in convex lensing are commonly used in the literature; our design combines two common refraction profiles in order to minimize lens aberration in the ``optical'' (ray) transport limit,
\begin{eqnarray}
\label{eqn:1} \bar{n}^{2}(y) & = & \frac{1}{\left[\bar{c}_{avg}(y)\right]^{2}} \nonumber \\
& = & \left(\beta \frac{\bar{c}_{0}^{2}}{1-\left(\alpha_{1} y\right)^{2}} + \left(1-\beta\right)\bar{c}_{0}^{2} \left(\cosh{\alpha_{2} y}\right)^{2}\right)^{-1} \\
\label{eqn:2} \alpha_{1} & = & (2/h)\sqrt{1-\left(\bar{c}_{0}/\bar{c}_{h}\right)^{2}} \\
\label{eqn:3} \alpha_{2} & = & (2/h)\operatorname{arccosh}\,(\bar{c}_{h}/\bar{c}_{0})\,
\end{eqnarray}
where $h$ is the lens height in the gradient direction, $\bar{c}_{0}=\bar{c}_{avg}(0)$ and $\bar{c}_{h}=\bar{c}_{avg}(h/2)$ are the extremal sound speeds obtained from the chosen range of $a_{y}/a_{x}$, and $\beta$ is a parameter that mixes the two GRIN profiles.  

An iterative ray tracing routine based on the eikonal approximation~\cite{Parazzoli2006JOSAB} was used to determine the optimal value of $\beta$ that minimizes the lens aberration.  The isotropic refraction gradient in Eq.~(\ref{eqn:1}) prescribes $\bar{c}_{avg}(y)$ and hence the lattice geometry $a_{y}/a_{x}$ for a given value of $\beta$; however, to optimize $\beta$ an eikonal approximation that accounts for the anisotropic sound speeds in the lattice was employed to improve the accuracy of the physical geometry.  The differential equation governing the eikonal function $\xi(\mathbf{x})$ in the presence of anisotropy can be derived for acoustic metafluids in a similar manner to the electromagnetic case,~\cite{Parazzoli2006JOSAB}
\begin{eqnarray}
\label{eqn:4} \bar{c}_{xx}^{2}\xi_{x}^{2}+\bar{c}_{yy}^{2}\xi_{y}^{2}-2\bar{c}_{xy}^{2}\xi_{x}\xi_{y} = 1
\end{eqnarray}
where $\xi_{[x,y]}=d\xi/d[x,y]$.  The off-diagonal sound speed $\bar{c}_{xy}$ can be calculated from the band structure or by considering multiple scattering effects.~\cite{Torrent2008NJPb}  The off-diagonal term was found to be negligible compared to the on-diagonal terms for each anisotropy ratio in our lattice.  

Figure~\ref{fig:2}(a) shows the ray paths derived using Eq.~(\ref{eqn:4}) for a GRIN lens insonified by a plane wave after optimizing the parameter $\beta$.  The lens has thickness $10a_{x}$, height $h=27.1a_{x}$, and lattice spacing $a_{x}=15.6$~mm; these parameters were also used to manufacture the physical lens.  The choice of $\beta=0.65$ produces a focal point at $x\simeq 1.2$~m from the center of the lens with minimal aberration of the ray trajectories.  The $\beta$-optimization routine starts with the average sound speed profile $\bar{c}_{avg}(y)$ as input and converts to the anisotropic phase speeds from Fig.~\ref{fig:1}(b).  The optimization routine also ensures that the gradient in the lattice spacing $a_y(y)$ is properly discretized to be consistent with the total height $h$ of the optimized GRIN profile.  

A schematic of the upper half of the $\beta$-optimized lens design is shown in Fig.~\ref{fig:2}(b). The full lens is symmetric about the $y$-axis.  A photograph of the assembled lens is shown in Fig.~\ref{fig:2}(c).  The lens is constructed of identical 1~m long hollow aluminum cylinders with air inside. Each cylindrical tube has a diameter of 12.5~mm and nominal wall thickness of 0.625~mm.  Rubber end caps are adhered to both ends of each cylinder to prevent water infiltration.  Additionally, $\sim 2.5$~cm of foam was inserted into the tube ends to damp the excitation of axial modes. The tubes are mounted on plexiglass end-plates into which holes with the designed lattice pattern are drilled. Two additional rows of cylinders with spacing $a_{y}=a_{x}$ are placed at the top and bottom of the lens to help mitigate the sharp transition where the edge of the lens ($\bar{c}_{avg}=1.36$) meets the open water ($\bar{c}_{avg}=1$).

\begin{figure*}[p]
\begin{center}
\includegraphics{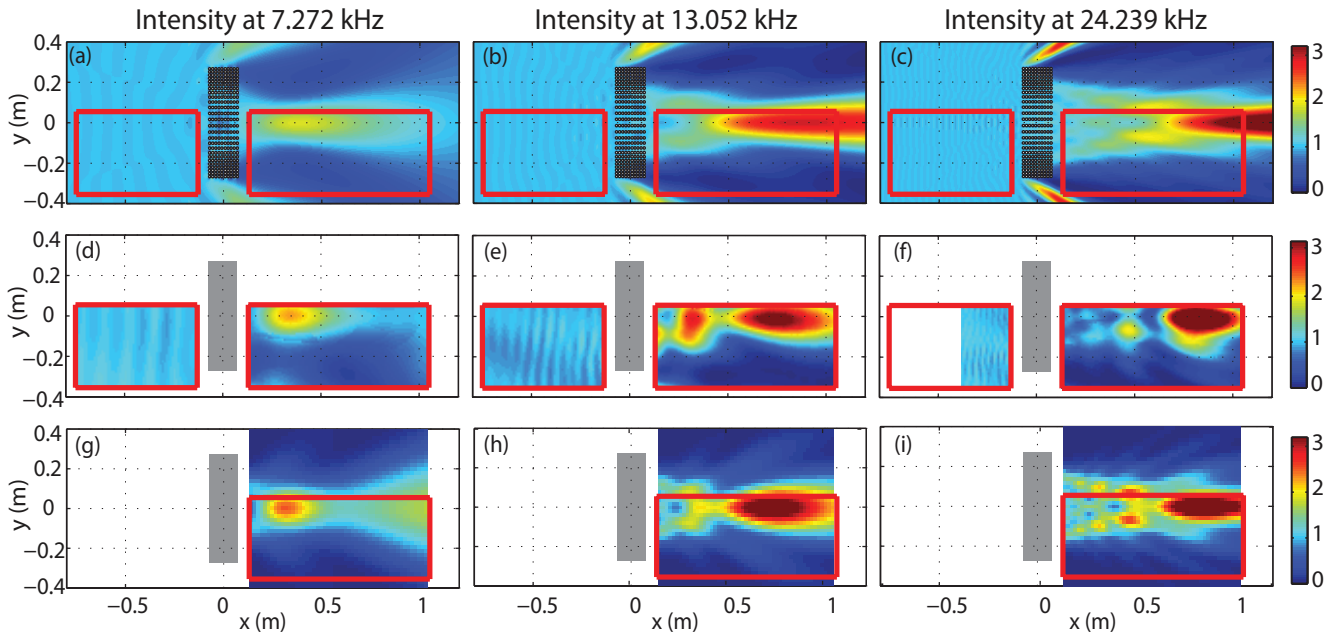}
\caption{Sound pressure intensity in the vicinity of the lens at $\omega a_{x}/2\pi c_{b}=0.077$ (7.27~kHz), 0.138 (13.05~kHz), and 0.256 (24.24~kHz) for a source location on the $x$-axis.  (a-c) Pressure intensity calculated using 2D-MST.  (d-f) Measured pressure intensity.  (g-i) Pressure intensity calculated using a 3D Rayleigh-Sommerfeld approximation.  Grey boxes show the position of the lens.  Red outlines indicate the experimentally mapped areas.  The reduced observational range of the reflected signal at 24.24 kHz is due to a limited overlap between the incident and reflected signals in the time domain at high frequency.}
\label{fig:3}
\end{center}
\end{figure*}

\begin{figure*}[p]
\begin{center}
\includegraphics {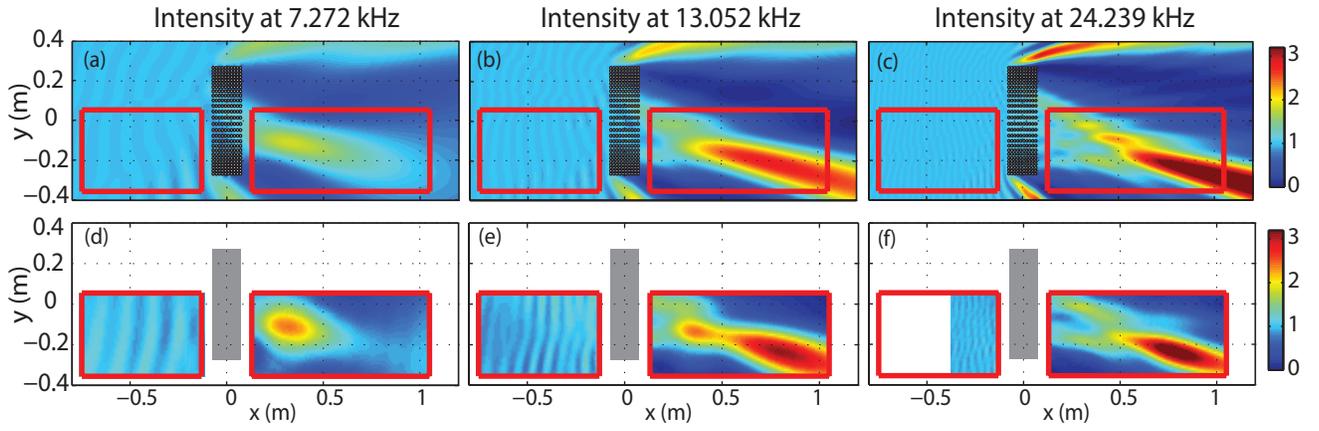}
\caption{Sound pressure intensity in the vicinity of the lens at $\omega a_{x}/2\pi c_{b}=0.077$ (7.27~kHz), 0.138 (13.05~kHz), and 0.256 (24.24~kHz) for a source positioned at a $15^{\circ}$~angle with respect to the $x$-axis.  (a-c) Pressure intensity calculated using 2D-MST.  (d-f) Measured pressure intensity.  Grey boxes show the position of the lens.  Red outlines indicate the experimentally mapped areas.}
\label{fig:4}
\end{center}
\end{figure*}

The lens was submerged at the center of a $6\times6\times4$~ m$^3$ water tank. Acoustic waves were produced by a 0.1~m diameter spherical source located in the plane that bisects the lens midpoint in the axial direction ($z$-axis).  In order to demonstrate the lens directionality, two in-plane source locations were considered at $(x,y) = (-1.6,0)$~m and $(x,y) = (-1.6,0.43)$~m corresponding to incident angles of $0^{\circ}$ and $15^{\circ}$ respectively from the lens central axis ($x$-axis). Wave propagation was measured using hydrophones at a sampling rate of 0.8~MHz.  Hydrophones were mounted on a three-axis translation positioning system to record measurements of the sound pressure levels, $P(x,y)$ and $P_{0}(x,y)$, in the presence and absence of the lens respectively. Measurements were taken at 10 mm increments. The sound pressure level was measured at individual frequencies by averaging over a 10-cycle pulse in the transmission region (as indicated in Figs.~\ref{fig:3} and \ref{fig:4}) and a 20-cycle pulse in the reflection region. The longer pulse cycle used in the reflection region ensured overlap between incident and reflected waves so that the measured total field could be properly compared with simulations. The length of the incident pulses were short enough to isolate reflections from the tank walls. 

Measurements were performed in the $xy$-plane perpendicular to the cylindrical axis of the lens over a broad range of frequencies $\omega a_{x}/2\pi c_{b}<0.3$.  Examples of the measured and numerically modeled pressure intensity $(|P|/|P_{0}|)^{2}$ in the vicinity of the lens are shown in Figs.~\ref{fig:3} and \ref{fig:4} for frequencies $\omega a_{x}/2\pi c_{b}=0.077$, 0.138, and 0.256.  In both the measured and predicted results, the pressure intensity is normalized (at each position) to that of the source intensity in the absence of the lens.  Given the symmetry of the lens about the $y$-axis, measurements were obtained over the bottom half of the scattering plane with a small overlap into the upper half-plane to detail the on-axis forward scattering pattern. The regions scanned in the measurement are also outlined in the upper panels of Figs.~\ref{fig:3} and~\ref{fig:4} for ease of comparison with the numerical predictions. Note that although the measurements at $\omega a_{x}/2\pi c_{b}=0.077$ and 0.256 lie within two of the resonance bands identified in Fig.~\ref{fig:1}, no obvious additional resonant features are observed in the intensity maps that can be attributed to these resonance bands.

The upper panels (a-c) of Figs.~\ref{fig:3} and~\ref{fig:4} show a prediction of the transmitted and reflected pressure intensity calculated using two-dimensional multiple scattering theory (2D-MST).~\cite{Martin2012PRB,Torrent2008NJPb}  The MST accounts for elastic scattering in the cylindrical tubes and assumes insonification by a cylindrical monopole source located at positions that match the experiment.  The experimentally measured acoustic intensities in panels (d-f) show good qualitative agreement with the MST-predicted results.  The reflected signal, $R=20\log_{10}[(P-P_{0})/P_{0})]$, was estimated relative to the incident amplitude $P_0$ using the sound pressure level $P$ measured in front of the incident face of the lens (regions to the left of the lens in Figs.~\ref{fig:3} and~\ref{fig:4}).  The measured reflection was at or below $13\%$ of the incident amplitude ($R\le-18$~dB) over the range of operational frequencies, which demonstrates significant transparency and is commensurate with recent Fresnel lens designs in air.~\cite{Moleron2014APL,Li2014NSR}  The measured reflection includes additional constructive interference compared to the MST due to diffraction from the finite aperture size in the axial direction.  

In the forward direction a focusing peak is observed that strengthens in intensity and moves out away from the lens with increasing frequency.  The frequency-dependence of the focusing peak is expected.  At very low frequencies the aperture is diffraction-limited; the wavelength is too long to adequately resolve the index gradient and there is also significant diffraction around the aperture.  As the frequency increases, the transmission approaches the geometric limit of ray-acoustics where a focusing peak would be located beyond the focal point predicted by the ray tracing in Fig.~\ref{fig:2}(a).  At the lowest example frequency, $\omega a_{x}/2\pi c_{b}=0.077$ [panels (a,d)], the lens is close to the diffraction limit with height-to-wavelength ratio $h/\lambda<2$.  As the frequency is further reduced the focusing peak intensity becomes significantly suppressed.  Therefore, an operational bandwidth can be identified ranging between $\omega a_{x}/2\pi c_{b}\approx0.05$--0.30, where the lower cutoff is constrained by the diffraction limit and the upper cutoff by the homogenization limit of the lattice.  

It is apparent in Figs.~\ref{fig:3} and~\ref{fig:4} that the 2D-MST does not accurately predict the precise location of the intensity maxima nor the small-scale, near-field interference features in the forward direction.  The discrepancy is due to the assumption of infinite cylinders in the 2D theory compared to the finite size of the lens aperture in the experiment.  The diffractive corrections arising from the finite size in the axial direction can be predicted using a three-dimensional (3D) Rayleigh-Sommerfeld integral.  The pressure at a position $(x,y,z)$ in front of an acoustic aperture located at $x=0$ can be calculated by integrating over the aperture,~\cite{born1999}
\begin{eqnarray}
\label{eqn:5} P(x,y,z) & = & \frac{x}{i\lambda} \iint\limits_{A} \tilde{P}(y',z') \frac{e^{ikr}}{r^2} dy' dz' \\
\label{eqn:6} \tilde{P}(y',z') & = & \psi(y',z')e^{i\phi(y',z')} \\
\label{eqn:7} r^{2} & = & x^{2}+\left(y-y'\right)^{2}+\left(z-z'\right)^{2}\,
\end{eqnarray}
where $\psi(y',z')$ and $\phi(y',z')$ are the real-valued acoustic amplitude and phase of the complex pressure $\tilde{P}(y',z')$ on the aperture facet, and $\lambda$ and $k$ are the wavelength and wave number respectively.  Here the integration aperture is simplified by the flat facet of our lens design, with the integration carried out over a simple rectangular area at the \emph{forward} face of the lens.

\begin{figure}[p]
\begin{center}
\includegraphics[width=13cm] {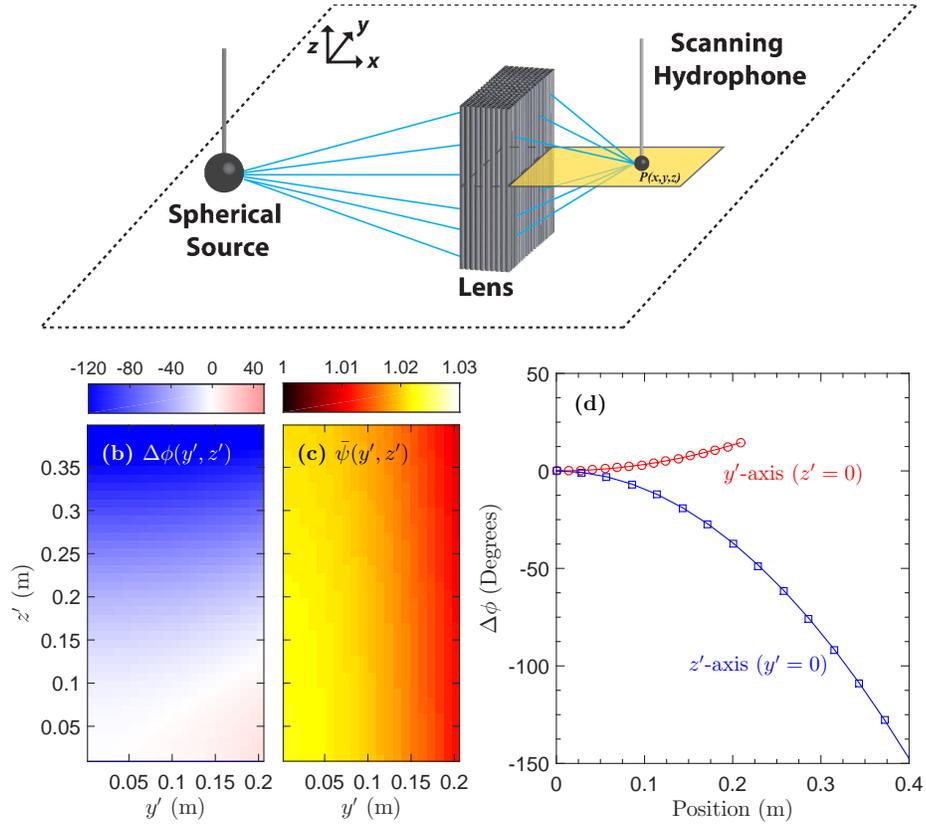}
\caption{(a) Schematic showing the experimental setup; blue lines indicate ray paths emerging from the spherical source and Rayleigh-Sommerfeld integration vectors from the forward facet of the lens to hydrophone locations in the measurement plane. (b) Change in phase $\Delta \phi(y^{\prime},z^{\prime})$ (degrees) over the forward facet (due to symmetry only the 1st quadrant is shown).  Blue regions represent a phase delay, red regions represent a phase advance.  (c) Normalized amplitude $\bar{\psi}(y^{\prime},z^{\prime})$ over the forward facet, where normalization is with respect to the amplitude of the spherical wave in the absence of the lens.  (d) Red and blue lines show the phase change $\Delta \phi$ plotted as a function of position along the $y^{\prime}$- and $z^{\prime}$-axes respectively.  Data points are calculated by advancing the phase along ray paths based on an eikonal approximation, lines are quadratic fits to the data.}
\label{fig:5}
\end{center}
\end{figure}

The pressure on the aperture facet $\tilde{P}(y',z')$ can be calculated using anisotropic ray-tracing under the eikonal approximation of Eq.~(\ref{eqn:4}).  A schematic in Figure~\ref{fig:5}(a) depicts the ray trajectories of wave-fronts that emerge from the spherical source and traverse the forward facet of the lens. The calculation iterates over a discrete set of rays that enter the \emph{rear} facet (i.e.~the incident face) at equally spaced intervals. The phase function $\phi(y',z')$ on the forward facet was determined by advancing the phase of each ray along the path lengths defined by $\xi(\mathbf{x})$ assuming a constant phase at the source. The effective path length is altered by the spatially-dependent local sound speed in the lens.  The phase difference $\Delta \phi=\phi(y',z')-\phi(0,0)$ relative to the phase at the lens central axis is shown in Fig.~\ref{fig:5}(b), where positive or negative values represent a phase advance or delay respectively.  The calculation is performed for a source position on the lens central axis.  Given the symmetry of the lens only one quadrant is shown.  Curve fits in Fig.~\ref{fig:5}(d) indicate that $\Delta \phi$ is approximately quadratic in both the $y$- and $z$-directions.  

The amplitude function $\psi(y',z')$ is inversely related to the square root of the local areal density of the rays.  This was approximated by calculating the average nearest-neighbor separation of rays intersecting the forward facet; a relative amplitude $\bar{\psi}(y',z')=\psi(y',z')/\psi(0,0)$ can then be estimated from the inverse of this nearest-neighbor separation.  Figure~\ref{fig:5}(c) shows the variation in $\bar{\psi}(y',z')$ after normalizing to the predicted amplitude in the absence of the lens.  It is clear that the lens refraction has minimal impact on the amplitude ($<3\%$) and that the diffractive correction is primarily influenced by the change in phase over the aperture.

Following a similar procedure as was used by Gao \emph{et.~al.},~\cite{Gao2012OE} Eq.~(\ref{eqn:5}) is evaluated numerically within the 2D plane of the measurement using quadratic approximations to $\phi(y',z')$ and $\psi(y',z')$.  Intensity maps produced by the Rayleigh-Sommerfeld integration are shown in Fig.~\ref{fig:3}(g-i) and demonstrate significantly improved quantitative agreement with the experiment.  As was the case for the measurement and MST modeling, the Rayleigh-Sommerfeld-based pressure intensities are normalized to the pressure intensity of the spherical source in the absence of the lens.  The 3D Rayleigh-Sommerfeld integration more accurately predicts both the location of the focal positions and their magnitude.  The measured magnification in sound pressure level ranges between 4.5--7.3~dB over the operational bandwidth.  The 2D-MST predicts magnifications that are as much as $40\%$ lower than the measured values, whereas the magnifications predicted by the Rayleigh-Sommerfeld integration agree to within $10\%$.  We emphasize that the 2D lattice only focuses along one axis ($y$-axis) resulting in lower magnifications than would be produced by axisymmetric 3D designs.~\cite{Moleron2014APL,Li2014NSR}  The magnification can be increased by extending our 2D design to three dimensions using tubes bent into a toroid configuration.~\cite{Sanchis2010APL,Chang2012NJP}  

We now place our results in context with other metafluid design concepts.  Traditional underwater and ultrasound applications have achieved an impedance-match by utilizing rubbers that have sound speeds less than water.  While rubbers have been used as components in transparent metafluid lattices,~\cite{Naify2014APL} there have been other recent advances in acoustic impedance-matched designs.  For example, negative-index complementary metamaterials have been proposed to significantly enhance the transparency through aberrating materials.~\cite{Shen2014PRX}  A class of metafluids based on the concept of space-coiling have also featured prominently in the literature.~\cite{Kock1949JASA,Liang2012PRL,Xie2013APL}  Space-coiled Fresnel lens designs have been reported that show significant focusing while utilizing a thin aperture compared to the wavelength.~\cite{Moleron2014APL,Li2014NSR}   Although the strength of the space-coiling design is the ability to significantly modify phase within a confined space, there is a drawback that it can only delay the phase over a lengthened path; similar to rubbers in water, the effective sound speed of these devices is less than the background fluid.  The space-coiling design has not yet been demonstrated in water, where a higher viscosity and lack of rigid boundary conditions may impose additional challenges.   

Our design extends the reach of aqueous transparent metafluids to include the option of \emph{phase advance} and decreased effective path length.  While not necessarily required for traditional lensing, there are metamaterial applications that require higher sound speeds compared to the propagation medium,~\cite{Maldovan2013Nat,Kadic2013RPP,Ren2010APL,Layman2011APL,Garcia2014PRB} the most prominent of which is scattering reduction.  The requirement of higher relative sound speeds features prominently in scattering reduction designs based on both coordinate transformation~\cite{Cummer2007NJP,Chen2007APL,Cummer2008PRL} and scattering cancellation.~\cite{Guild2012PRB,Martin2012APL}  The realization of a high-sound speed, transparent GRIN lens represents an important proof-of-concept: it demonstrates that phase-advance metafluids can be constructed with significant tunability in both sound speed and impedance, including the option of matched impedance.  Given that the concept of core-shell lattice components has now been extended to also include core modification,~\cite{Titovich2014JASA} a wide range of GRIN and anisotropic designs should be achievable in water based on metafluid lattices.

[Work is supported by the Office of Naval Research.]

\end{document}